\documentclass[12pt,aps,prb,preprint]{revtex4}

\usepackage{amsmath}    % need for subequations
\usepackage{graphicx}   % for figures

\begin{document}
\newcommand{\ab}{\v{a}} 
\newcommand{\ai}{\^{a}} 
\newcommand{\ib}{\^{\i}} 
\newcommand{\tb}{\c{t}} 
\newcommand{\st}{\c{s}}
\newcommand{\Ab}{\v{A}} 
\newcommand{\Ai}{\^{A}} 
\newcommand{\Ib}{\^{I}} 
\newcommand{\Tb}{\c{T}}
\newcommand{\St}{\c{S}}

\newcommand{\muv}{\boldsymbol{\mu}}
\newcommand{\mc}{{\mathcal M}}
\newcommand{\pc}{{\mathcal P}}
 \newcommand{\mv}{\boldsymbol{m}}
\newcommand{\pv}{\boldsymbol{p}}
\newcommand{\tv}{\boldsymbol{t}}
\def\msf{\hbox{{\sf M}}}
\def\msft{\boldsymbol{{\sf M}}}
\def\psf{\hbox{{\sf P}}}
\def\psft{\boldsymbol{{\sf P}}}
\def\Nsf{\hbox{{\sf N}}}
\def\Nsft{\boldsymbol{{\sf N}}}
\def\Tsf{\hbox{{\sf T}}}
\def\Tsft{\boldsymbol{{\sf T}}}
\def\Asf{\hbox{{\sf A}}}
\def\Asft{\boldsymbol{{\sf A}}}
\def\Bsf{\hbox{{\sf B}}}
\def\Bsft{\boldsymbol{{\sf B}}}
\def\Lsf{\hbox{{\sf L}}}
\def\Lsft{\boldsymbol{{\sf L}}}
\def\Ssf{\hbox{{\sf S}}}
\def\Ssft{\boldsymbol{{\sf S}}}
\def\Mtens{\bi{M}}
\def\msfsim{\boldsymbol{{\sf M}}_{\scriptstyle\rm(sym)}}
\newcommand{\mcsim}{ {\sf M}_{ {\scriptstyle \rm {(sym)} } i_1\dots i_n}}
\newcommand{\mcs}{ {\sf M}_{ {\scriptstyle \rm {(sym)} } i_1i_2i_3}}

\newcommand{\beqan}{\begin{eqnarray*}}
\newcommand{\eeqan}{\end{eqnarray*}}
\newcommand{\beqa}{\begin{eqnarray}}
\newcommand{\eeqa}{\end{eqnarray}}

 \newcommand{\suml}{\sum\limits}
 \newcommand{\sumd}{\suml_{\mathcal D}}
\newcommand{\intl}{\int\limits}
\newcommand{\rvec}{\boldsymbol{r}}
\newcommand{\xvec}{\boldsymbol{x}}
\newcommand{\xivec}{\boldsymbol{\xi}}
\newcommand{\Avec}{\boldsymbol{A}}
\newcommand{\Rvec}{\boldsymbol{R}}
\newcommand{\Evec}{\boldsymbol{E}}
\newcommand{\Bvec}{\boldsymbol{B}}
\newcommand{\Svec}{\boldsymbol{S}}
\newcommand{\avec}{\boldsymbol{a}}
\newcommand{\nablav}{\boldsymbol{\nabla}}
\newcommand{\nuvec}{\boldsymbol{\nu}}
\newcommand{\bvec}{\boldsymbol{\beta}}
\newcommand{\vvec}{\boldsymbol{v}}
\newcommand{\jvec}{\boldsymbol{J}}
\newcommand{\nvec}{\boldsymbol{n}}
\newcommand{\pvec}{\boldsymbol{p}}
\newcommand{\mvec}{\boldsymbol{m}}
\newcommand{\evec}{\boldsymbol{e}}
\newcommand{\eps}{\varepsilon}
\newcommand{\la}{\lambda}
\newcommand{\rad}{\mbox{\footnotesize rad}}
\newcommand{\scr}{\scriptstyle}
\newcommand{\latens}{\boldsymbol{\Lambda}}
\newcommand{\pitens}{\boldsymbol{\Pi}}
\newcommand{\cm}{{\cal M}}
\newcommand{\cp}{{\cal P}}
\newcommand{\beq}{\begin{equation}}
\newcommand{\eeq}{\end{equation}}
\newcommand{\ptens}{\boldsymbol{\sf{P}}}
\newcommand{\Ptens}{\boldsymbol{P}}
\newcommand{\Ttens}{\boldsymbol{\sf{T}}}
\newcommand{\Ntens}{\boldsymbol{\sf{N}}}
\newcommand{\Ncal}{\boldsymbol{{\cal N}}}
\newcommand{\Atens}{\boldsymbol{\sf{A}}}
\newcommand{\Btens}{\boldsymbol{\sf{B}}}
\newcommand{\dom}{\mathcal{D}}
\newcommand{\al}{\alpha}
\newcommand{\sym}{\scriptstyle \rm{(sym)}}
\newcommand{\Tcal}{\boldsymbol{{\mathcal T}}}
\newcommand{\Nmc}{{\mathcal N}}
\renewcommand{\d}{\partial}
\def\rmi{{\rm i}}
\def\rme{\hbox{\rm e}}
\def\rmd{\hbox{\rm d}}
\newcommand{\ct}{\mbox{\Huge{.}}}
\newcommand{\Laop}{\boldsymbol{\Lambda}}
\newcommand{\Ssfs}{{\scriptstyle \Ssft^{(n)}}}
\newcommand{\Lsfs}{{\scriptstyle \Lsft^{(n)}}}
\newcommand{\psfr}{\widetilde{\psf}}
\newcommand{\msfr}{\widetilde{\msf}}
\newcommand{\msftr}{\widetilde{\msft}}
\newcommand{\psftr}{\widetilde{\psft}}
\newcommand{\qdot}{\stackrel{\cdot\cdot\cdot\cdot}}
\newcommand{\tdot}{\stackrel{\cdot\cdot\cdot}}
\newcommand{\eref}{(\ref}
\newcommand{\bsy}{\boldsymbol}
\newcommand{\dotj}{\boldsymbol{\dot{J}}}
\newcommand{\psfs}{{\sf P}}
\newcommand{\msfs}{{\sf M}}
\newcommand{\Fvec}{\boldsymbol{F}}
\newcommand{\Qvec}{\boldsymbol{Q}}
\newcommand{\bmu}{\bsy{\mu}}
\newcommand{\bpi}{\bsy{\pi}}
\newcommand{\lasf}{{\sf\Lambda}}
\newcommand{\lasft}{\bf{\sf \Lambda}}
\newcommand{\pisf}{{\sf \Pi}}
\newcommand{\pisft}{\bf{\sf \Pi}}
\newcommand{\gamsf}{{\sf \Gamma}}
\newcommand{\gamsft}{\bf{\sf \Gamma}}
\newcommand{\Scal}{{\mathcal S}}

\title{Some alternatives for calculating multipole expansions of the electromagnetic radiation field}
\author{C. Vrejoiu and R. Zus}
% \altaffiliation[Also at ]{home.}  %  optional
 \affiliation{University of Bucharest, Department of Physics, Bucharest, Romania} 
 \email{vrejoiu@fizica.unibuc.ro;roxana.zus@fizica.unibuc.ro}   %optional
%\author{Jan Tobochnik}
%\affiliation{Kalamazoo College, Department of Physics, Kalamazoo,
% 49007}
\date{\today}
\begin{abstract}
We discuss the multipolar expansion of the electromagnetic field with an emphasis on the radiated field. We investigate if the employment of Jefimenko's equations brings a new insight into the calculation of the radiation field. We show that the affirmation is valid if one finds an interesting example in which inverting the order between spatial derivatives and integration is not allowed.  Further, we consider the generalization of the multipolar expansion of the power radiated by a confined system of charges and currents to a higher arbitrary order. 
\end{abstract}

\maketitle
\section{Introduction}\label{sec:intro}
Despite the successful and long history of the electromagnetic field theory, there are several topics open to new theoretical and pedagogical contributions. One of them concerns the formalism of multipole expansion of the field, in general, and of the radiated one, in particular. Another issue is related to the importance of Jefimenko's equations in the study of such problems.  Motivated by recent publications on the topic, we discuss some features of this type of problems. 
\par The multipole expansion of the electromagnetic field in Cartesian coordinates is exposed in electrodynamics textbooks, as the well-known Refs. \cite{Landau} and \cite{Jackson}.
 Ordinarily, these expansions are calculated only in the first two or three orders, the higher-orders being considered too complicated. As Jackson writes in his textbook, {\it the labor involved in manipulating terms in the expansion of the vector potential becomes increasingly prohibitive as the expansion is extended beyond the electric quadrupole terms} (see Ref. \cite{Jackson}, pp 415-416). 
 For this reason and due to the applicability only in the long-wavelength range, another treatment, based on the spherical tensors and the solutions of Helmholtz equation is preferred. This alternative has also a larger domain of applications. Actually, starting from the results obtained employing this calculation technique, the reader can verify what effort is involved when returning to the multipole Cartesian moments which offer a higher physical transparency (see Ref. \cite{RV}). A relatively recent textbook \cite{Raab} and a paper \cite{Melo}, the last  related to the importance of Jefimenko's equations for expressing the electric and magnetic field when discussing the radiation theory, brought our attention on a very hard  formalism employed for the calculation of even the first three or four terms of the expansion series. Though there are some prescriptions in the literature \cite{Thorne}, \cite{Damour}, \cite{cvJPA} for calculating higher-order terms of the multipole series based on a simple algebraic formalism of tensorial analysis, it seems that there is some reticence in using this last technique. For this reason, one of the aims of the present paper is to show how one can hide, as much as possible, the higher-order tensors behind some vectors, reducing the calculation technique to the formalism of an ordinary vectorial algebra or analysis. 
 \par Another aim of the paper is to investigate if the use of Jefimenko's equations brings, as sometimes presented in the literature, a new insight in the calculation of the radiation field. We show that unless one finds an example where the spatial derivative and the integral operations can not be inverted, it is not always a necessary complication. As already mentioned, we also use the opportunity to explain some advanced features of the multipolar expansions in the field radiation theory.
 \par We start in section \ref{sec:gen} by shortly presenting the notation convention we use and by giving a general formalism for handling multipolar expansions in Cartesian coordinates.  In section \ref{sec:radiation}, we derive the radiated electric and magnetic field  without using the retarded potentials, while in  section \ref{sec:jefimenko} 
 we present characteristics of the calculation for the radiation field when employing Jefimenko's equations.   The advantages and disadvantages of different approaches are analyzed. In section \ref{sec:aspects}, we further discuss some features of the calculation for the radiated power, with an emphasis on the $4-th$ order approximation in $d/\la\,<\,1$. Finally, in section \ref{sec:formulae}, we give the guidelines for the general tensorial calculus of the electric and magnetic moments. The last section is reserved for conclusions.

\section{General formalism}\label{sec:gen}
We write  Maxwell equations with a notation independent of the unit system  (``system free'' Maxwell equations):
\beqa\label{Maxwell}
\nablav\times\Bvec&=&\frac{\mu_0}{\al}\left(\jvec+\eps_0\frac{\d\Evec}{\d t}\right),\;\;\nablav\times\Evec=-\frac{1}{\al}\frac{\d\Bvec}{\d t},\nonumber\\
\nablav\cdot\Bvec&=&0,\;\;\;\nablav\cdot\Evec=\frac{1}{\eps_0}\rho\ ,
\eeqa
where $\eps_0,\,\mu_0,\,\al$ are proportional factors  depending on the system of units and are satisfying the equation
\beqa\label{relc}
\frac{\al^2}{\eps_0\mu_0}=c^2.
\eeqa
$c$ is the vacuum light speed.
Maxwell equations written in SI units are obtained from equations \eref{Maxwell}) for  $\al=1$ and the SI values of $\eps_0,\;\mu_0$. For the Gauss system of units, $\al=c,\,\eps_0=1/4\pi,\,\mu_0=4\pi$. With this notation, Jefimenko's  equations  can be written as \cite{Jackson}  
\beqa\label{JefB}
\Bvec(\rvec,t)=\frac{\mu_0}{4\pi\al}\intl_{\dom}\nablav\times\frac{[\jvec]}{R}\,\rmd^3x',
\eeqa
\beqa\label{JefE}
\Evec(\rvec,t)=-\frac{1}{4\pi\eps_0}\intl_{\dom}\nablav\frac{[\rho]}{R}\,\rmd^3x'-\frac{\mu_0}{4\pi\al^2}\intl_{\dom}\frac{[\dot{\jvec}]}{R}\,\rmd^3x',
\eeqa
where $\Rvec=\rvec-\rvec',\;[f]=f(\rvec',t-R/c)$ and the origin $O$ of Cartesian coordinates is  in the domain $\dom$. The support of charge and current distribution is supposed included in $\dom$. If in equations \eref{JefB}) and \eref{JefE}) the order of the derivative and the integral is inverted, one obtains the well-known relations between fields and potentials:
\beqa\label{F-pot}
\Bvec=\nablav\times\Avec,\;\,\Evec=-\nablav\Phi-\frac{1}{\al}\frac{\d\Avec}{\d t}\ ,
\eeqa
with the retarded potentials
\beqa\label{potret}
\Avec(\rvec,t)=\frac{\mu_0}{4\pi\al}\intl_{\dom}\frac{[\jvec]}{R}\,\rmd^3x',\;\; \Phi(\rvec,t)=\frac{1}{4\pi\eps_0}\intl_{\dom }\frac{[\rho]}{R}\,\rmd^3x'\ .
\eeqa
\par In \cite{Melo}, the authors  derive  the multipole expansion of the radiation field from equations \eref{JefB}) and \eref{JefE})  claiming  to give an original  demonstration specific for Jefimenko's equations, without  employing the retarded potentials. We should agree with this claim if at least some calculation of the authors is different from  those employing the potential multipole expansions which are  generally used in literature. In the following, we search for a difference between the calculation presented in Ref. \cite{Melo} and the standard one making use of potentials. The goal of the exposition below is  to inform on  some results regarding multipolar expansion in Cartesian coordinates, too. 
\par Let us derive the multipolar expansion  of the field $\Bvec$ given by equation \eref{JefB}), and written explicitly as:
\beqa\label{B1}
\Bvec(\rvec,t)= \frac{\mu_0}{4\pi\al}\evec_i\eps_{ijk}\intl_{\dom}\d_j\,\frac{[J_k]}{R}\,\rmd^3x',
\eeqa
where $\evec_i$ are the unit vectors of the Cartesian axes. 
Writing the integral from equation \eref{B1}) as
\beqa\label{B2}
\intl_{\dom}\d_j\,\frac{[J_k]}{R}\,\rmd^3x'=\intl_{\dom}\,\d_j\left(\frac{J_k(\xivec,t-\frac{R}{c})}{R}\right)_{\xivec=\rvec'}\,\rmd^3x',
\eeqa
we obtain the multipolar expansion of the magnetic field about $O$ as function of $\rvec'$ using the Taylor series of the integrand:
\beqan
\d_j\frac{J_k(\xivec,t-\frac{R}{c})}{R}=\suml_{n\ge 0}\frac{(-1)^n}{n!}x'_{i_1}\dots x'_{i_n}\,\d_{i_1}\dots\d_{i_n}\,\d_j\left(\frac{1}{r}J_k(\xivec,t-\frac{r}{c})\right).
\eeqan
Equation \eref{B1}) can now be  expressed as
\beqa\label{B3}
\Bvec(\rvec,t)=\frac{\mu_0}{4\pi\al}\evec_i\,\eps_{ijk}\intl_{\dom}\,\suml_{n\ge 0}\frac{(-1)^n}{n!}x'_{i_1}\dots x'_{i_n}
\d_j\d_{i_1}\dots \d_{i_n}\,\left(\frac{1}{r}\,[J_k]_0\right)\,\rmd^3x'
\eeqa
where we employed the notation $[f]_0=f(\rvec',\,t-r/c)$.
\par We assume one is allowed to invert  orders of operations in equation \eref{B3}) and to write:
\beqa\label{B4-0}
\!\!\!\!\!\!\!\!\Bvec(\rvec,t)=\frac{\mu_0}{4\pi\al}\evec_i\eps_{ijk}
\suml_{n\ge 0}\frac{(-1)^n}{n!}\d_j\d_{i_1}\dots\d_{i_n}\left(\frac{1}{r}\intl_{\dom}x'_{i_1}\dots x'_{i_n}[J_k]_0\right)\,\rmd^3x'.
\eeqa
Equation \eref{B4-0}) represents the curl of the multipolar expansion of the vector potential $\Avec$. Thus one can perform firstly the multipolar expansion of this potential. It is the usual procedure. 
\par No matter what  procedure is  employed, a constant in the calculation is the presence of a vector $\bsy{a}(\rvec,t;\bsy{\zeta},n)$ defined by the Cartesian components:
\beqa\label{arn}
\!\!\!\!a_k(\rvec,t;\bsy{\zeta},n)=\zeta_{i_1}\dots \zeta_{i_n}
\left(\frac{1}{r}\intl_{\dom}x'_{i_1}\dots x'_{i_n}\,J_k(\rvec',t)\,\rmd^3x'\right).
\eeqa
Here, $\zeta$ can be either an operator or a number.
Generalizing to the dynamic case an algorithm used in  \cite{Castell} for the  magnetostatic field, we introduce in equation \eref{arn}) the consequence of the continuity equation, written for $t_0=t-r/c$:
\beqan
J_k(\rvec',t_0)=\nablav'\big(x'_k\jvec(\rvec',t_0)\big)+x'_k\dot{\rho}(\rvec',t_0).
\eeqan
We obtain
\beqa\label{arn1}
\!\!\!\!\!\!\!\!&~&a_k(\rvec,t_0;\bsy{\zeta},n)=\zeta_{i_1}\dots \zeta_{i_n}\frac{1}{r}\intl_{\dom}x'_{i_1}\dots x'_{i_n}\,\nablav'\left(x'_k[\jvec]_0\right)\,\rmd^3x'
+\zeta_{i_1}\dots \zeta_{i_n}\frac{1}{r}\dot{\psfs}_{i_1\dots i_n\,k}(t_0),
\eeqa
where the Cartesian components of the $n-th$ electric moment of the given charge distribution:
\beqa\label{pel}
\psfs_{i_1\dots i_n}(t)=\intl_{\dom}x'_{i_1}\dots x'_{i_n}\,\rho(\rvec',t)\,\rmd^3x'
\eeqa
are introduced. Let us define, for simplifying the notation, the vector 
\beqa\label{vpe}
\bsy{\mathcal P}(\rvec,t;\bsy{\zeta},n)&=&\evec_k\,\zeta_{i_1}\dots \zeta_{i_{n-1}}\,\frac{\psfs_{i_1\dots i_{n-1}\,k}(t)}{r}.
\eeqa
Performing partial integration and taking into account that $\jvec$ vanishes  on the surface $\d\dom$,  equation \eref{arn1}) can be written and processed as follows:
\beqa\label{arn2}
\!\!&~&a_k(\rvec,t_0;\bsy{\zeta},n)- \dot{\mathcal P}_k(\rvec, t;\bsy{\zeta},n+1)
=-\zeta_{i_1}\dots\zeta_{i_n}\frac{1}{r}\intl_{\dom} x'_k\,[\jvec]_0\cdot\nablav'(x'_{i_1}\dots x'_{i_n})\,\rmd^3x'\nonumber\\
\!\!\!\!&=&-n\,\zeta_{i_1}\dots\zeta_{i_n}\frac{1}{r}\intl_{\dom}x'_{i_1}\dots x'_{i_{n-1}}\,x'_k\,[J_{i_n}]_0\,\rmd^3x'\nonumber\\
&=&-n\,\zeta_{i_1}\dots\zeta_{i_n}
\frac{1}{r}\intl_{\dom}x'_{i_1}\dots x'_{i_{n-1}}\big(x'_k\,[J_{i_n}]_0-x'_{i_n}\,[J_k]_0\big)\,\rmd^3x'
-n a_k(\rvec,t_0;\bsy{\zeta},n)\nonumber\\
\!\!\!\!\!\!\!\!\!\!&=&-n\,\eps_{ki_nq}\,\zeta_{i_n}\,\zeta_{i_1}\dots \zeta_{i_{n-1}}
\frac{1}{r}\intl_{\dom}x'_{i_1}\dots x'_{i_{n-1}}\big(\rvec'\times[\jvec]_0\big)_q\,\rmd^3x'
-n a_k(\rvec,t_0;\bsy{\zeta},n).
\eeqa
Introducing the $n-th$ order magnetic moment, as in Ref. \cite{Castell}, by its Cartesian components
\vspace{-0.3cm}
\beqa\label{M}
\msfs_{i_1\dots i_n}(t)
=\frac{n}{(n+1)\al}\intl_{\dom}x'_{i_1}\dots x'_{i_{n-1}}\left(\rvec'\times\jvec(\rvec',t)\right)_{i_n}\,\rmd^3x',
\eeqa
equation \eref{arn2}) becomes
\beqa\label{arn3}
\!\!\!\!\!\!\!\!\!\!\!\!a_k(\rvec,t_0;\bsy{\zeta},n)=-\al\,\eps_{ki_nq}\zeta_{i_n}\,\zeta_{i_1}\dots \zeta_{i_{n-1}}\frac{\msfs_{i_1\dots i_{n-1}q}(t_0)}{r}
+\frac{1}{n+1}\dot{\mathcal P}_k(\rvec,t_0;\bsy{\zeta},n+1).
\eeqa
Similarly to equation \eref{vpe}), we introduce the vector
\beqa\label{vpm}
\bsy{\mathcal M}(\rvec,t;\bsy{\zeta},n)&=&\evec_k\,\zeta_{i_1}\dots \zeta_{i_{n-1}}\,\frac{\msfs_{i_1\dots i_{n-1}\,k}(t)}{r}\ ,
\eeqa
writing finally equation \eref{arn3}) as
\beqa\label{arn4}
\bsy{a}(\rvec,t_0;\bsy{\zeta},n)=-\al\bsy{\zeta}\times\bsy{\mathcal M}(\rvec,t_0;\bsy{\zeta},n)
+\frac{1}{n+1}\dot{\bsy{\mathcal P}}(\rvec,t_0;\bsy{\zeta},n+1)\ .
\eeqa
With this result, the magnetic field from equation \eref{B4-0}) can be expressed with the help of the vectors $\bsy{\mathcal M}$ and $\bsy{\mathcal P}$:
\beqan
&~&\Bvec(\rvec,t)=\frac{\mu_0}{4\pi\al}\evec_i\eps_{ijk}\suml_{n\ge 0}\frac{(-1)^n}{n!}\,\d_j\,a_k(\rvec,t_0;\nablav,n)\\
&~&=\frac{\mu_0}{4\pi}\suml_{n\ge 1}\frac{(-1)^{n-1}}{n!}\nablav\times\big(\nablav\times\bsy{\mathcal M}(\rvec,t_0;\nablav,n)\big)
+\frac{\mu_0}{4\pi\al}\suml_{n\ge 0}\frac{(-1)^n}{(n+1)!}\nablav\times\dot{\bsy{\mathcal P}}(\rvec,t_0;\nablav,n+1),
\eeqan
or, by a change of the summation index in the second sum,
\beqa\label{B5}
\Bvec(\rvec,t)=\nablav\times\frac{\mu_0}{4\pi}\suml_{n\ge 1}\frac{(-1)^{n-1}}{n!}\left(\nablav\times\bsy{\mathcal M}(\rvec,t_0;\nablav,n)
+\frac{1}{\al}\dot{\bsy{\mathcal P}}(\rvec,t_0;\nablav,n)\right).
\eeqa
From the last expression one has  no problem in identifying  the multipolar expansion of the vector potential $\Avec$:
\beqa\label{Adezv}
\Avec(\rvec,t)=\frac{\mu_0}{4\pi}\suml_{n\ge 1}\frac{(-1)^{n-1}}{n!}\left(\nablav\times\bsy{\mathcal M}(\rvec,t_0;\nablav,n)+\frac{1}{\al}\dot{\bsy{\mathcal P}}(\rvec,t_0;\nablav,n)\right).
\eeqa
\par The calculation for the electric field can be performed in a similar manner. One obtains:
\beqa\label{E1}
\!\!\!\!\!\!\!\!\!\!\!\!\!\!\Evec(\rvec,t)&=&-\frac{1}{4\pi\eps_0}\suml_{n\ge 0}\frac{(-1)^n}{n!}\nablav\big(\nablav\cdot\bsy{\mathcal P}(\rvec,t_0;\nablav,n)\big)\nonumber\\
\!\!\!\!\!\!\!\!\!\!\!\!\!\!&~&-\frac{\mu_0}{4\pi\al}\suml_{n\ge 1}\frac{(-1)^{n-1}}{n!}\left(\nablav\times\dot{\bsy{\mathcal M}}(\rvec,t_0;\nablav,n)
+\frac{1}{\al}\ddot{\bsy{\mathcal P}}(\rvec,t_0;\nablav,n)\right)
\eeqa
Comparing equations \eref{F-pot}) with \eref{Adezv}) and \eref{E1}), we single out  the  multipole expansion of the potential $\Phi$:
\beqa\label{Phi}
\Phi(\rvec,t)=\frac{1}{4\pi\eps_0}\suml_{n\ge 0}\frac{(-1)^n}{n!}\nablav\cdot\bsy{\mathcal P}(\rvec,t_0;\nablav,n).
\eeqa
\section{Radiation field}\label{sec:radiation}
\par For calculating the radiation field it is sufficient retaining only  terms of order $1/r$ and $1/r^2$ for $r\to \infty$. In most  textbooks one retains only the terms of order $1/r$, the goal being, usually, only the derivation of the radiated energy or of the linear momentum. Actually, when the goal is the complete definition of the radiation field, one must be able to derive all  transferring properties, including the angular momentum loss. These are, in fact, minimal conditions for defining a physical system. In the last case,  the terms of order $1/r^2$ are also necessary (see Ref. \cite{Landau} -Problem 2 at the end of Section 72, and also Refs. \cite{RV},\cite{cvcs}). 
Although the aim of the present paper is different, we also give the formula for introducing terms of order $1/r^2$ required for the evaluation of the angular momentum loss. The terms of the orders $1/r$ and $1/r^2$ are selected making use of formula \cite{cvcs}:
\beqa\label{deriv}
\!\!\!\!\!&~&\d_{i_1}\dots \d_{i_n}\left(\frac{f(t_0)}{r}\right)=\frac{1}{r}\,\frac{(-1)^n}{c^n}\nu_{i_1}\dots \nu_{i_n}\frac{\d^nf(t_0)}{\d t^n}\nonumber\\
\!\!\!\!\!&~&+\frac{(-1)^n}{c^{n-1}r^2}\left(D_n\,\nu_{i_1}\dots\nu_{i_n}
-\nu_{\{i_1\dots i_{n-2}}\delta_{i_{n-1}i_n\}}\right)\frac{\d^{n-1}f(t_0)}{\d t^{n-1}}.
\eeqa
 
Again $t_0=t-r/c$ and $\nu_i=x_i/r$. By $A_{\{i_1\dots i_n\}}$ we understand 
 the sum over all the permutations of the symbols $i_q$ that give distinct terms. The coefficients $D_n$ are defined by the recurrence relations:
\beqa\label{Dn}
D_n=D_{n-1}+n,\, D_0=0.
\eeqa
The formula from equation \eref{deriv}) can be easily proven by recurrence.
\par In fact,  formula \eref{deriv}) represents the sum of the terms corresponding to   $l=0$ and $1$ of a general formula 
\beqa\label{derivgen}
\d_{i_1}\dots \d_{i_n}\,\frac{f(t_0)}{r}=\suml^{n}_{l=0}C^{(n,l)}_{i_1\dots i_n}(\nuvec)\frac{1}{r^{l+1}}\frac{\d^{n-l}}{\d t^{n-l}}f(t_0).
\eeqa
In this equation, $C^{(n,l)}_{i_1\dots i_n}(\nuvec)$ are symmetric coefficients expressed as linear combinations of products of  components  $\nu_i$ and Kronecker symbols $\delta_{i_qi_p}$,where $i,\,i_q,\,i_p= i_1\dots i_n$.
 \par Considering equations \eref{vpe}) and \eref{vpm}), one can see that $\bsy{\mathcal P}(\rvec,t_0;\bsy{\zeta},n)$ and $\bsy{\mathcal M}(\rvec,t_0;\bsy{\zeta},n)$ are solutions of the homogeneous wave equation for $r\ne 0$. Consequently, one can apply the formula \eref{deriv}) for these quantities. Let us consider the multiple derivative  as, for example, 
 \beqa\label{dM}
&~& \d_{j_1}\dots\d_{j_m}\bsy{\mathcal M}(\rvec,t_0;\nablav,n)=\frac{(-1)^{n+m-1}}{c^{n+m-1}}\nu_{j_1}\dots\nu_{j_m}
 \nu_{i_1}\dots\nu_{i_{n-1}}\,\evec_k
 \frac{\d^{n+m-1}}{\d t^{n+m-1}}\frac{\msfs_{i_1\dots i_{n-1}\,k}(t_0)}{r}\nonumber\\
 &~&\,+\,{\mathcal O}(1/r^2)=\frac{(-1)^{n+m-1}}{c^{n+m-1}}\nu_{j_1}\dots\nu_{j_m}\frac{\d t^{n+m-1}}{\d t^{n+m-1}}\bsy{\mathcal M}(\rvec,t_0;\nuvec,n)
 +{\mathcal O}(1/r^2)
 \eeqa
and similarly for $\bsy{\mathcal P}(\rvec,t_0;\nablav,n)$. Using equation \eref{dM}) and the equivalent relation for 
$\bsy{\mathcal P}(\rvec,t_0;\nablav,n)$ in equation \eref{B5}), we obtain the first approximation of the multipolar expansion  of radiated magnetic field, which is sufficient for calculating the radiated energy and the linear momentum:
\beqa\label{Brad1}
&~&\Bvec_{rad}(\rvec,t)=\frac{\mu_0}{4\pi}\suml_{n\ge 1}\frac{1}{n!c^{n+1}}\frac{\d^{n+1}}{\d t^{n+1}}
\left(\nuvec\times\big(\nuvec\times\bsy{\mathcal M}(\rvec,t_0;\nuvec,n)\big)
-\frac{c}{\al}\nuvec\times\bsy{\mathcal P}(\rvec,t_0;\nuvec,n)\right)\\
&~&=\frac{\mu_0}{4\pi}\suml_{n\ge 1}\frac{1}{n!c^{n+1}}\frac{\d^{n+1}}{\d t^{n+1}}\left(\big(\nuvec\cdot\bsy{\mathcal M}(\rvec,t_0;\nuvec,n)\big)\,\nuvec
-\bsy{\mathcal M}(\rvec,t_0;\nuvec,n)-\frac{c}{\al}\nuvec\times\bsy{\mathcal P}(\rvec,t_0;\nuvec,n)\right).\nonumber
\eeqa
For the field $\Evec_{rad}$,  we obtain
\beqa\label{Erad} 
%\begin{multline}
\Evec_{rad}(\rvec,t)&=&\frac{1}{4\pi\eps_0}\suml_{n\ge 1}\frac{1}{n!c^{n+1}}\frac{\d^{n+1}}{\d t^{n+1}}
\big(\nuvec\cdot\bsy{\mathcal P}(\rvec,t_0;\nuvec,n)\,\nuvec-\bsy{\mathcal P}(\rvec,t_0;\nuvec,n)\nonumber\\
 &~&\;\;\;\;\;\;\;\;\;\;\;\;\;\;\;\;\;\;\;\;\;\;\;\;\;\;\;\;\;\;\;\;\;\;\;\;\;\;\;\;\;\;\;\;\;\;\;\;+\frac{\al}{c}\nuvec\times\bsy{\mathcal M}(\rvec,t_0;\nuvec,n)\big).
%\end{multline}
\eeqa
We made use of equation \eref{relc}).
From equations \eref{Adezv}) and \eref{Phi}), using equation \eref{deriv}), we get easily the expansions of the radiation field potentials:
\beqa\label{dArad}
\Avec_{rad}(\rvec,t)=\frac{\mu_0}{4\pi}\suml_{n\ge 1}\frac{1}{n!c^n}\frac{\d^n}{\d t^n}\left(\bsy{\mathcal M}(\rvec,t_0;\nuvec,n)\times \nuvec
+\frac{c}{\al}\bsy{\mathcal P}(\rvec,t_0;\nuvec,n)\right)
\eeqa
and
\beqa\label{dPhirad}
\!\!\!\!\!\!\!\Phi_{rad}(\rvec,t)=\frac{1}{4\pi\eps_0}\suml_{n\ge 1}\frac{1}{n!c^n}\frac{\d^n}{\d t^n}\,\nuvec\cdot\bsy{\mathcal P}(\rvec,t_0;\nuvec,n).
\eeqa
Given the above expressions, one can verify that the relations between fields and potentials are:
\beqa\label{Frad-Prad}
\Bvec_{rad}&=&\frac{1}{c}\big(\dot{\Avec}_{rad}\times\nuvec\big),\;\;\;
\Evec_{rad}=\frac{1}{\al}\big(\dot{\Avec}_{rad}\times\nuvec\big)\times\nuvec =\frac{c}{\al}\Bvec_{rad}\times\nuvec.
\eeqa
These parts (proportional to $1/r$) from the radiated electric and magnetic fields  are purely transverse fields, satisfying the properties (see also \cite{cvcs}):
\beqa\label{transv}
\nuvec\cdot\Evec_{rad}=0,\;\;\nuvec\cdot\Bvec_{rad}=0,\;\;
\eps_0\vert\Evec_{rad}\vert^2=\frac{1}{\mu_0}\vert\Bvec_{rad}\vert^2.
\eeqa

\section{Is really the radiation field calculation from Jefimenko's equations a new insight in the radiation theory ?}
\label{sec:jefimenko}
In Ref. \cite{Landau}, calculating the radiation field (in the first approximation), the retarded potentials are approximated by the formulae (66.1) and (66.2) of this reference, written here with the ``system free'' notation:
\beqa\label{L1}
\Phi(\rvec,t)\approx\frac{1}{4\pi\eps_0\,r}\intl_{\dom}\rho(\rvec',t_0+\frac{1}{c}\nuvec\cdot\rvec')\,\rmd^3x',\;\;
\Avec(\rvec,t)\approx\frac{\mu_0}{4\pi\al\,r}\intl_{\dom}\jvec(\rvec',t_0+\frac{1}{c}\nuvec\cdot\rvec')\,\rmd^3x'.
\eeqa
As suggested in Ref. \cite{Landau} (page 184, first footnote), introducing these expressions of the radiated potentials and retaining only the terms of order $1/r$, one obtains the relations from equation \eref{Frad-Prad}) ( i.e. equation (66.3) from Ref. \cite{Landau}).
 In this calculation, the derivative operators must be introduced in the integral and so, the proof is indeed realized directly for the fields $\Evec$ and $\Bvec$. For $\Bvec$, using the relations
\beqan
\nablav\times\frac{\jvec(\rvec',t_0+\frac{\nuvec\cdot\rvec'}{c})}{r}
=(\nablav\frac{1}{r})\times\jvec+\frac{1}{r}\nablav(t_0+\frac{\nuvec\cdot\rvec'}{c})\times \dot{\jvec}
\eeqan
and
\beqa\label{aprox}
\nablav\frac{1}{r}={\mathcal O}(\frac{1}{r^2}),\;\;\;
\nablav(t_0+\frac{1}{c}\nuvec\cdot\rvec')=-\frac{1}{c}\nuvec+{\mathcal O}(\frac{1}{r}),
\eeqa
we can write
\beqa\label{B6}
\widetilde{\Bvec}(\rvec,t)&=&\frac{\mu_0}{4\pi\al}\intl_{\dom}\nablav\times\frac{\jvec(\rvec',t_0+\frac{\nvec\cdot\rvec'}{c})}{r}\,\rmd^3x'\nonumber\\
&=&-\frac{\mu_0}{4\pi\al c}\,\frac{1}{r}\nuvec\times\intl_{\dom}\dot{\jvec}(\rvec',t_0+\frac{1}{c}\nuvec\cdot\rvec')\,\rmd^3x'
+{\mathcal O}(\frac{1}{r^2}).
\eeqa
From this last equation we can extract the term of  order  $1/r$ which represents the radiated field (precisely, the first approximation of this field):
\beqa\label{B7}
&~& ~\\
&~&\Bvec_{rad}(\rvec,t)=-\frac{\mu_0}{4\pi\al c\,r}\,\nuvec\times\intl_{\dom}\dot{\jvec}(\rvec',t_0+\frac{1}c\nuvec\cdot\rvec')\,\rmd^3x'.\nonumber
\eeqa
This is, in fact,  equation (28) from Ref. \cite{Melo} and we can consider, by examining  Ref. \cite{Landau}, that this result was obtained long time ago  from the expression \eref{JefB}) of the field $\Bvec$ (see Ref. \cite{cvedp} and Ref.\cite{cvdn-arx}). Look also in  Ref. \cite{Melo} to see the usage of Jefimenko's equation for calculating $\Bvec_{rad}$. 
\par Introducing also the approximate expression $\widetilde{\Evec}$ starting from equation \eref{JefE}), we obtain
\beqa\label{Eaprox}
\widetilde{\Evec}(\rvec,t)&=&-\frac{\mu_0}{4\pi \al^2\,r}\intl_{\dom}\left(c^2\nablav \rho(\rvec',t_0+\frac{1}{c}\nuvec\cdot\rvec')+\dot{\jvec}(\rvec',t_0+\frac{1}{c}\nuvec\cdot\rvec')\right)\,\rmd^3x'+{\mathcal O}(\frac{1}{r^2})\nonumber\\
&=&\frac{\mu_0}{4\pi\al^2\,r}\intl_{\dom}\left(c\,\dot{\rho}(\rvec',t_0+\frac{1}{c}\nuvec\cdot\rvec')\,\nuvec-\dot{\jvec}(\rvec',t_0+\frac{1}{c}\nuvec\cdot\rvec')\right)\,+\,{\mathcal O}(\frac{1}{r^2}).
\eeqa
Using the continuity equation written in the point $\rvec'$ at the retarded time $t-R/c$, 
\beqa\label{ec-aprox}
[\dot{\rho}]=-[\nabla'\cdot\jvec(\rvec',t')]_{t'=t-R/c}=-\nablav'\cdot[\jvec]+[\dot{\jvec}]\cdot\nabla'(t-R/c),
\eeqa
it results
\beqa\label{ec-aprox2}
\dot{\rho}(\rvec',t_0+\frac{1}{c}\nuvec\cdot\rvec')=-\nablav'\cdot\jvec(\rvec',t_0+\frac{1}{c}\nuvec\cdot\rvec')+\frac{1}{c}\nuvec\cdot\dot{\jvec}(\rvec',t_0+\frac{1}{c}\nuvec\cdot\rvec')+{\mathcal O}(\frac{1}{r}).
\eeqa
The first term from the right-hand side of the last equation gives no contribution to the integral from equation \eref{Eaprox}) and, after a simple algebraic calculation, we obtain the expression of $\Evec_{rad}$ from equation  \eref{Frad-Prad})
(see also equations (66.3) from Ref. \cite{Landau} and equation(30) from Ref. \cite{Melo}).
\par We point out that equation \eref{ec-aprox}) or, generally, the relation between the space derivative of a retarded quantity and the retarded value of the space derivative of the same quantity,  should be  well-known for each student from a class of electrodynamics since when writing the retarded potentials as solutions of the wave equation, it is necessary to verify the Lorenz condition. The verification can be realized in a direct calculation, a good exercise for a student. Only  in this way one can be convinced that the retarded solutions are indeed electromagnetic potentials. For this goal, equation \eref{ec-aprox}) is indispensable since the Lorenz condition appears as a consequence of the continuity equation.
\par Based on the above  results, we can quickly obtain  the multipole expansion of the radiation field. Considering the adequate Taylor series for the integrand in equation \eref{B6}), we write
\beqa\label{B8a}
\!\!\!\!\!\!\!\Bvec_{rad}(\rvec,t)&=&-\frac{\mu_0}{4\pi\al c}\,\frac{1}{r}\,\nuvec\times\intl_{\dom}\left(\dot{\jvec}(\xivec,t_0+\frac{1}{c}\nuvec\cdot\rvec')\right)_{\xivec=\rvec'}\,\rmd^3x'\nonumber\\
&=&-\frac{\mu_0}{4\pi\al}\nuvec\times\suml_{n\ge 0}\frac{1}{n!c^{n+1}}\frac{\d^n}{\d t^n}\left(\nu_{i_1}\dots \nu_{i_n}\,\frac{1}{r}\intl_{\dom}x'_{i_1}\dots x'_{i_n}\dot{\jvec}(\rvec',t_0)\,\rmd^3x'
\right)\nonumber\\
&=&-\frac{\mu_0}{4\pi\al}\suml_{n\ge 0}\frac{1}{n!c^{n+1}}\frac{\d^{n+1}}{\d t^{n+1}}\,\nuvec\times \bsy{a}(\rvec,t_0;\nuvec,n).
\eeqa
The vector $\bsy{a}$  was defined in equation \eref{arn}). After introducing its expression as a function of the vectors associated to the electric and magnetic moment, equation \eref{arn4}), the radiated magnetic field becomes 
\beqa\label{B8}
\Bvec_{rad}(\rvec,t)&=&\frac{\mu_0}{4\pi}\suml_{n\ge 1}\frac{1}{n!c^{n+1}}\frac{\d^{n+1}}{\d t^{n+1}}\nuvec\times\big(\nuvec\times\bsy{\mathcal M}(\rvec,t_0;\nuvec,n)\big)\nonumber\\
&~& \;\;\;\;\;\;\;\;\;\;\;\;\;\;\;\;\;\;-\frac{\mu_0}{4\pi\al}\suml_{n\ge 0}\frac{1}{(n+1)!c^{n+1}}\frac{\d^{n+2}}{\d t^{n+2}}\,\nuvec\times\bsy{\mathcal P}(\rvec,t_0;\nuvec,n+1).
\eeqa
The final step is  changing $n\to n-1$ in the second sum of equation \eref{B8}). We obtain again equation \eref{Brad1}). 
\par We can say  that  the result \eref{B8}), which is the multipole expansion of the radiation field,  was determined  via Jefimenko's equations.
 \par Alternatively,  after reaching the expression given by equation \eref{B3}), we can perform the multipolar expansion of the radiation field in such a manner that we can also say  that it is obtained via Jefimenko's equations. We extract from the integrand in equation \eref{B3}) the terms of the order $1/r$ (and $1/r^2$  for a complete definition of this field):
 \beqa\label{B9}
 \Bvec_{rad}(\rvec,t)&=&-\,\frac{\mu_0}{4\pi\al}\evec_i\eps_{ijk}\, \nu_j\,\nu_{i_1}\dots\nu_{i_n}\frac{1}{r}\intl_{\dom}\suml_{n\ge 0}\frac{1}{n!c^{n+1}}\,\,x'_{i_1}\dots x'_{i_n}\frac{\d^{n+1}}{\d t^{n+1}}[J_k]_0\,\rmd^3x'\nonumber\\
&=&-\,\frac{\mu_0}{4\pi\al}\evec_i\eps_{ijk}\suml_{n\ge 0}\frac{1}{n!c^{n+1}}\nu_j\,\frac{\d^{n+1}}{\d t^{n+1}}a_k(\rvec',t_0;\nuvec,n),
 \eeqa
arriving, as expected, to equation \eref{B8a}). 

\par We remind the reader that for obtaining the results of equations \eref{B5}) and \eref{E1}) we admitted the commutation of the derivative with respect to the spatial coordinates,  with the series expansion and the integral on the domain $\dom$. In the present section, for the case of the radiation field, we avoid the inversion of the derivative with the integral operation. Regarding the commutation with the Taylor expansion, we consider that such an operation cannot be avoided as long as we want to emphasize the multipolar moments. 
\par As a short extension in the argumentation of the utility of Jefimenko's equations, we show how one can avoid the commutation between derivative and integral for an arbitrary point in the exterior of the domain $\dom$. Indeed,  using equation \eref{derivgen}) in equation \eref{B3}), we can write the magnetic field with the help of time derivatives
 \beqa\label{B3'}
 \Bvec(\rvec,t)=\frac{\mu_0}{4\pi\al}\evec_i\eps_{ijk}\suml_{n\ge 0}\frac{(-1)^n}{n!}\,\suml^{n+1}_{l=0}\frac{1}{r^{l+1}}\,\frac{\d^{n+1-l}}{\d t^{n+1-l}}\,\widetilde{a}_{jk}(\rvec,t_0;\nuvec,n).
 \eeqa
Here, 
\beqan
\widetilde{a}_{jk}(\rvec,t;\nuvec,n)=C^{(n+1,l)}_{j\,i_1\dots i_n}\,\intl_{\dom}x'_{i_1}\dots x'_{i_n}[J_k]_0\,\rmd^3x'.
\eeqan
Similar to the derivation of equation \eref{arn3}) from equation \eref{arn}), we get:
\beqan
\widetilde{a}_{jk}(\rvec,t_0;\nuvec,n)=-\al\,\eps_{ki_nq}\,C^{(n+1,\,l)}_{j\,i_1\dots i_n}\,\msfs_{i_1\dots i_{n-1}\,q}(t_0)
+\frac{1}{n+1}\,C^{(n+1,\,l)}_{j\,i_1\dots i_n}\dot{\psfs}_{i_1\dots i_n\,k}.
\eeqan
Inserting this equation in equation \eref{B3'}), we obtain
\beqa\label{B4'}
\Bvec(\rvec,t)&=&\frac{\mu_0}{4\pi}\suml_{n\ge 1}\frac{(-1)^{n-1}}{n!}\,\evec_i\eps_{ijk}\suml^{n+1}_{l=0}\frac{\eps_{ki_nq}}{r^{l+1}}C^{(n+1,l)}_{j\,i_1\dots i_n}\frac{\d^{n+1-l}}{\d t^{n+1-l}}\msfs_{i_1\dots i_{n-1}\,q}\nonumber\\
&~&\;\;\;\;\;\;\;\;\;\;\;\;+\frac{\mu_0}{4\pi\al}\suml_{n\ge 0}\frac{(-1)^n}{(n+1)!}\evec_i\eps_{ijk}\suml^{n+1}_{l=0}\frac{1}{r^{l+1}}C^{(n+1,l)}_{j\,i_1\dots i_n}
\frac{\d^{n+2-l}}{\d t^{n+2-l}}\psfs_{i_1\dots i_n\,k}
\eeqa
As anticipated, it represents the multipolar expansion of the magnetic field for any point in the exterior of the domain $\dom$ and it is obtained without inverting the spatial derivative with the integral operation. For the approximation of the radiated field (in $1/r$), one can easily verify that writing equation \eref{B4'}) for $l=0$ and taking 
\beqan
C^{(n+1,0)}_{j\,i_1\dots i_n}=\frac{(-1)^{n+1}}{c^{n+1}}\nu_j\nu_{i_1}\dots \nu_{i_n},
\eeqan
we get equation \eref{Brad1}).
\par In conclusion, employing Jefimenko's equation in the radiation theory could bring a new insight only if the inversion of the spatial derivative and the integral operation is not allowed. We admit we were unable to find an interesting example where an inversion is not permitted, at least for generalized distributions. However, it might be possible to find such examples, and, in this case, the indispensable character of Jefimenko's equations would be obvious. Otherwise, for the regular cases, it appears as an unnecessary complication.

\section{Some features of the radiated power calculation}\label{sec:aspects}

The Poynting vector is
\beqa\label{S}
\bsy{S}=\frac{\al}{\mu_0}\big(\Evec\times\Bvec\big)=\frac{c}{\mu_0}\vert\Bvec_{rad}\vert^2\,\nuvec+{\mathcal O}(1/r^3)=
\eps_0\,c\vert\Evec_{rad}\vert^2\,\nuvec + {\mathcal O}(1/r^3).
\eeqa
The total radiated power may be written as the limit of a surface  integral on a sphere  centered in $O$, of radius $r$,  for $r\to \infty$. Let us express the energy current in the radiation approximation corresponding to the sphere of radius $r$ at the time $t$:
\beqa\label{Sflux}
N(\bsy{S},\Sigma_r;t)=\oint\limits_{\Sigma_r}r^2\,\nuvec\cdot\bsy{S}(\rvec,t)\,\rmd\Omega(\nuvec)=\frac{c}{\mu_0}\oint\limits_{\Sigma_r}r^2\vert\Bvec_{rad}(\rvec,t)\vert^2\,\rmd\Omega(\nuvec)+{\mathcal O}(\frac{1}{r}).
\eeqa
The quantity $N(\bsy{S},\Sigma_r;t)\,\rmd t$ represents, therefore,  the energy which crosses the sphere $\Sigma_r$ in the time interval $(t,\,t+\rmd t)$ and is determined by the values of the multipolar moments of the source in the interval $(t-r/c, t+\rmd t-r/c)$. For large but finite $r$, the integrand from equation \eref{Sflux}) can be enployed for drawing conclusions on the electric charge distribution at the retarded time from observations on the angular distribution of radiation. 
Since $\rmd t=\rmd t_0$, we can say that $N(\bsy{S},\Sigma_r;t)\,\rmd t$  is the part of the energy emitted by source in the given time interval which contributes to the energy intensity corresponding to $\Sigma_r$ at the time $t$.  As one can see from equation \eref{Brad1}), if in  equation \eref{Sflux}) we put $t$ instead  $t_0$, then the quantity $\displaystyle\lim_{r\to\infty} N(\bsy{S},\Sigma_r;t)\,\rmd t$ represents that part of the energy emitted by the source  which contributes to the radiated energy or, shortly, radiated by the source. The situation changes when the support of the source depends on time and, in particular, for the radiation of a moving point-like source  (see \cite{Landau}, \& 73). In conclusion, the  energy emitted by the source  is characterized by the intensity 
\beqa\label{Irad}
{\mathcal I}=\lim_{r\to\infty}N(\bsy{S},\Sigma_r;t)=\frac{cr^2}{\mu_0}\int\vert\Bvec_{rad}(\rvec,t)\vert^2\,\rmd\Omega(\nuvec)=
\frac{4\pi c}{\mu_0}\langle r^2\vert\Bvec_{rad}\vert^2\rangle ,
\eeqa
where 
\beqan
\langle f\rangle=\frac{1}{4\pi}\int f(\nuvec)\,\rmd\Omega(\nuvec).
\eeqan
 $\Bvec_{rad}(\rvec,t)$ is given by equation \eref{Brad1}) substituting the retarded time $t_0$ by $t$. Since in equation \eref{Irad}) the factor $r^2$ is simplified  by the factors $1/r$ included in the definition of   ${\mathcal M}$ and ${\mathcal P}$  we should  rather  introduce  the quantities  $\bsy{\mu}$ and $\bsy{\pi}$:
\beqa\label{mupi}
\bsy{\mu}(t;\nuvec,n)=r\,\bsy{{\mathcal M}}(\rvec,t;\nuvec,n),\;\;\;\bsy{\pi}(t;\nuvec,n)=r\,\bsy{{\mathcal P}}(\rvec,t;\nuvec,n).
\eeqa
The expression for  the radiation intensity ${\mathcal I}$ is
\beqa\label{I0}
{\mathcal I}(t)=\frac{\al^2}{4\pi\eps_0c^3}\suml_{n,m\ge 1}&~&\frac{1}{n!m!\,c^{n+m}}\frac{\d^{n+1}}{\d t^{n+1}}\left\langle\big(\nuvec\cdot\bmu(t;\nuvec,n)\big)\,\nuvec-\bmu(t;\nuvec,n)-\frac{c}{\al}\,\nuvec\times\bpi(t;\nuvec,n)\right\rangle\nonumber\\
&~&\frac{\d^{m+1}}{\d t^{m+1}}\left\langle\big(\nuvec\cdot\bmu(t;\nuvec,m)\big)\,\nuvec-\bmu(t;\nuvec,m)-\frac{c}{\al}\,\nuvec\times\bpi(t;\nuvec,m)\right\rangle.
\eeqa
Employing  the notation $f_{,\,n}=\d^nf(\dots,t)/\d t^n$, and specifying only the argument $n$ in $\bmu$ and $\bpi$ when there is not a case of confusion, we can write
\beqa\label{I1}
{\mathcal I}(t)&=&\frac{\al^2}{4\pi\eps_0c^3}\suml_{n,m\ge 1}\frac{1}{n!m!\,c^{n+m}}\nonumber\\
&~&\left\langle-\big(\nuvec\cdot\bmu_{,n+1}(n)\big)
\big(\nuvec\cdot\bmu_{,m+1}(m)\big)\right.\nonumber\\
&~&\left.+\bmu_{,n+1}(n)\cdot\bmu_{,m+1}(m)+\frac{c}{\al}\bmu_{,n+1}(n)\cdot\left(\nuvec\times\bpi_{,m+1}(m\right))\right.\nonumber\\
  &~& +\left.\frac{c}{\al}\left(\nuvec\times\bpi_{,n+1}(n)\right)\cdot\bmu_{,m+1}(m)
  +\frac{c^2}{\al^2} \left\{\bpi_{,n+1}(n)\cdot\bpi_{,m+1}(m)\right.\right.\nonumber\\
 &~&\left.\left. -\big(\nuvec\cdot\bpi_{n+1}(n)\big)\big(\nuvec\cdot\bpi_{,m+1}(m)\big)\right\}  \right\rangle .
\eeqa     

One can calculate the averaged quantities from the last equation using  formula \cite{Thorne}:
\beqa\label{numed}
\langle\nu_{i_1}\dots\nu_{i_n}\rangle=\left\{\begin{array}{c}0,\;\;\;\;\;\;\;\;\;\;\;\;\;\;\;\;\;\;\;\;\;\;\;\;\;\;n=2k+1,\\
\frac{1}{(2k+1)!!}\,\delta_{\{i_1i_2}\dots\delta_{i_{n-1}i_n\}},\;\;\;\;\;\;\;\;\;\;n=2k,\;\;\;\;\;k=0,1,\dots\end{array}\right.
\eeqa
 In equation \eref{I1})  all terms containing an odd number of factors $\nu$ vanish and  we can retain only terms with an  even number of these factors. We point out that $\bmu(n)$ or $\bpi(n)$ contain $n-1$ factors $\nu$ (see equations \eref{vpe}), \eref{vpm}) and \eref{mupi})).
\par We have to calculate expressions as, for example,
\beqan
\langle\big(\nuvec\cdot\bmu(t;\nuvec,n)\big)\big(\nuvec\cdot\bmu(t;\nuvec,m)\big)\rangle=\langle\nu_i\,\nu_{i_1}\dots\nu_{n-1}\,\nu_j\,\nu_{j_1}\dots\nu_{j_{m-1}}\rangle \;\msfs_{i_1\dots i_{n-1}\,i}\,\,\msfs_{j_1\dots j_{m-1}\,j}.
\eeqan
This is cumbersome even for the first  approximations.
\par  Before discussing and applying the above results, we stress an essential issue for the existence of a precise approximation criteria when a finite number of terms in equation \eref{tpr}) is retained. Let us consider the source from $\dom$ being a system of $N$ point-like electric charges $q_1\dots q_N$. Therefore,
\beqa\label{rjq}
\rho(\rvec,t)=\suml^N_{i=1}q_i\,\delta\big(\rvec-\rvec^{(i)}(t)\big),\;\;\;
\jvec(\rvec,t)=\suml^N_{i=1}\,q_i\dot{\rvec}^{(i)}(t)\delta\big(\rvec-\rvec^{(i)}(t)\big),
\eeqa
where $\rvec^{(i)}(t)$ represents the position vector of the particle $i$. In this case,
\beqa\label{Pq}
\psfs_{i_1\dots i_n}(t)=\suml^N_{i=1}q_i\,x^{(i)}_{i_1}(t)\dots x^{(i)}_{i_n}(t)
\eeqa
and
\beqa\label{Mq}
\msfs_{i_1\dots i_n}(t)=\suml^N_{i=1}q_ix^{(i)}_{i_1}(t)\dots x^{(i)}_{i_{n-1}}(t)\big(\rvec^{(i)}\times\dot{\rvec}^{(i)}\big)_{i_n}(t).
\eeqa
Let us suppose the particles oscillating with a pulsation $\omega=2\pi c/\la$, i.e.
\beqa\label{osc}
\rvec^{(i)}(t)=\rvec^{(i)}_0\rme^{\rmi\omega t},\;\;\;\dot{\rvec}^{(i)}(t)=\rmi\omega\,\rvec^{(i)}_0\rme^{\rmi\omega t}=\frac{2\pi \rmi c}{\la}\,\rvec^{(i)}_0\rme^{\rmi\omega t}.
\eeqa
(For the general case of $N$ wave lengths $\la_i,\;\;i=1\dots N$,  in the following, we will understand by $\la$ the shortest of them.) Denoting by $d$ the linear dimension of the domain $\dom$, we have
%\beqan
$\vert\rvec^{(i)}_0\vert\,\le d.$
%\eeqan
 Considering the radiation in the case of a long wave-length, $\la\,>\,d$, we introduce the parameter 
 \beqa\label{d/la}
\zeta=\frac{d}{\la}\,<\,1.
\eeqa
For the amplitudes of the source, we identify the  orders of magnitude:
\beqa\label{omr}
\frac{\vert\rvec^{(i)}\vert}{\la}\,\lesssim \zeta,\;\;\;\vert\dot{\rvec}^{(i)}\vert\sim \frac{\vert\rvec^{(i)}\vert}{\la}\,\lesssim \zeta.
\eeqa
Obviously, the same relations can be written for any $x^{(k)}_i,\;\dot{x}^{(k)}_i$. For the time derivatives of the tensors 
$\psft^{(n)}$ and $\msft^{(n)}$  we can  conclude that
\beqa\label{ompm}
\frac{\d^k}{\d t^k}\psft^{(n)}\sim \left\{\begin{array}{c}\zeta^n,\;\;\;\;k\ge n\\
\zeta^k,\;\;\;\;k\le n\end{array}\right. ,\;\;\;\;\;\;
\frac{\d^k}{\d t^k}\msft^{(n)}\sim \left\{\begin{array}{c}\zeta^{n+1},\;\;\;\;k\ge n\\
\zeta^{k+1},\;\;\;\;k\le n\end{array}\right. .
\eeqa

%***************************************
With the approximation criteria considered above, we start by calculating the terms from equation \eref{I1}) up to the $4-th$ order in the parameter $\zeta$.
 For the electric and magnetic dipolar moments we use the usual notation $\pvec$ and $\mvec$, respectively. 
  We  select from equation \eref{I1}) all nonvanishing terms for $(n,m)=(1,1)$: 
\beqa\label{I11}
{\mathcal I}_{11}=
\frac{\al^2}{4\pi\eps_0c^5}\left(-\langle\nu_i\nu_j\rangle\ddot{m}_i\ddot{m}_j+\ddot{\mvec}^2+\frac{c^2}{\al^2}\left(\ddot{\pvec}^2-\langle\nu_i\nu_j\rangle\ddot{p}_i\ddot{p}_j\right)\right\rangle.
\eeqa
Inserting the result for the even combinations  $\langle\nu_i\nu_j\rangle$, we obtain the well-known expression for the radiation in the dipolar approximation:
\beqa\label{I11b}
{\mathcal I}_{11}=\frac{1}{6\pi\eps_0c^3}\left(\ddot{\pvec}^2+\frac{\al^2}{c^2}\ddot{\mvec}^2\right).
\eeqa
\par Now let us  consider in equation \eref{I1}) the terms with $(n,m)=(1,2)$ and $(2,1)$,  discarding the vanishing ones :
\beqa\label{I12}
{\mathcal I}_{12}+{\mathcal I}_{21}=\frac{\al}{8\pi\eps_0c^5}\left\langle\ddot{\bmu}(1)\cdot\big(\nuvec\times\tdot{\bpi}(2)\big)+\big(\nuvec\times\ddot{\bpi}(1)\big)\cdot\tdot{\bmu}(2)\right\rangle\ .
\eeqa
The first term, written explicitly with omission of  dots representing the time derivatives of no significance for the tensorial relations, is
\beqan\left\langle\bmu(1)\cdot\big(\nuvec\times\bpi(2)\big)\right\rangle=\left\langle m_i\big(\nuvec\times\bpi(2)\big)_i\right\rangle=\eps_{ijk} m_i\langle\nu_j\nu_q\rangle\psfs_{qk}=\frac{1}{3}\eps_{ijk}\delta_{jq}m_i\psfs_{qk}=0\ ,
\eeqan
because of the symmetry of the electric quadrupole moment. For the second term,
\beqan
 \left\langle\big(\nuvec\times\bpi(1)\big)\cdot\bmu(2)\right\rangle=\left\langle\eps_{ijk}\nu_jp_k\nu_q\msfs_{qi}\right\rangle=\frac{1}{3}p_k\eps_{kij}\msfs_{ji}\ ,
 \eeqan
 and finally,
 \beqa\label{I12c}
 {\mathcal I}_{12}+{\mathcal I}_{21}=-\frac{\al}{12\pi\eps_0c^5}\ddot{p}_k\,\eps_{kij}\,\tdot{\msfs}_{ij}. 
 \eeqa
\par Writing equation \eref{I1}) for $(n,m)=(2,2)$, we discard the terms of order $6$  in $\zeta$  as, for example, $\tdot{\bmu}(2)\cdot\tdot{\bmu}(2)$. In this approximation,
\beqa\label{I22}
{\mathcal I}_{22}&=&\frac{1}{16\pi\eps_0c^5}\left\langle \tdot{\bpi}^2(2)-\big(\nuvec\cdot\tdot{\bpi}(2)\big)^2\right\rangle+
{\mathcal  O}(\zeta^6)\nonumber\\
&=&\frac{1}{16\pi\eps_0c^5}\left(\langle\nu_i\nu_j\rangle\tdot{\psf}_{ik}\tdot{\psf}_{jk}-\langle\nu_i\nu_l\nu_j\nu_q\rangle\tdot{\psfs}_{li}\tdot{\psfs}_{qj}\right)+{\mathcal  O}(\zeta^6)\nonumber\\
&=&\frac{1}{16\pi\eps_0c^5}\left( \frac{1}{3}\delta_{ij}\tdot{\psfs}_{ik}\tdot{\psfs}_{jk}-\frac{1}{15}\delta_{\{il}\delta_{jq\}}\tdot{\psf}_{li}\tdot{\psfs}_{qj} \right)+{\mathcal  O}(\zeta^6)\nonumber\\
&=&\frac{1}{16\pi\eps_0c^5}\left( \frac{1}{3}\tdot{\psfs}_{ij}\tdot{\psfs}_{ij}-\frac{1}{15}\tdot{\psfs}_{ii} \tdot{\psfs}_{jj}-\frac{2}{15}\tdot{\psfs}_{ij}\tdot{\psfs}_{ij}\right)+{\mathcal  O}(\zeta^6)\nonumber\\
&=&\frac{1}{80\pi\eps_0c^5}\left(\tdot{\psfs}_{ij}\tdot{\psfs}_{ij}-\frac{1}{3}\big(\tdot{\psfs}_{ii}\big)^2\right)
+{\mathcal  O}(\zeta^6)\ .
\eeqa
Terms of order $6$  in $\zeta$ exist also for $(m,n)=(1,3)$ or $(3,1)$. Retaining only terms of order $4$,  we get:
\beqa\label{I13}
{\mathcal I}_{13}+{\mathcal I}_{31}&=&\frac{1}{12\pi\eps_0 c^5}\left\langle \ddot{\bpi}(1)\cdot\dot{\tdot{\bpi}}(3)-\big(\nuvec\cdot\ddot{\bpi}(1)\big)\big(\nuvec\cdot\dot{\tdot{\bpi}}(3)\big)\right\rangle+{\mathcal O}(\zeta^6)\nonumber\\
&=&\frac{1}{12\pi\eps_0 c^5}\left(\langle\nu_l\nu_qp_i\rangle\dot{\tdot{\psfs}}_{lqi}-\langle\nu_i\nu_j\nu_l\nu_q\rangle\ddot{p}_i\dot{\tdot{\psfs}}_{lqi}\right)+{\mathcal O}(\zeta^6)\nonumber\\
&=&\frac{1}{3}\delta_{lq}\ddot{p}_i\dot{\tdot{\psfs}}_{lqi}-\frac{1}{15}\delta_{\{ij}\delta_{lq\}}\ddot{p}_i\dot{\tdot{\psfs}}_{lqj}\nonumber+{\mathcal O}(\zeta^6)\\
&=&\frac{1}{90\pi\eps_0c^5}\,\ddot{p}_i\,\dot{\tdot{\psfs}}_{qqi}+{\mathcal O}(\zeta^6)\ .
\eeqa
Adding up  equations \eref{I11b}), \eref{I12c}), \eref{I22}) and \eref{I13}), we obtain the $4-th$ order approximation of the total radiated power:
\beqa\label{TRP}
{\mathcal  I}_{(4)}=\frac{1}{4\pi\eps_0c^3}\left(\frac{2}{3}\ddot{\pvec}^2+\frac{2}{3}\frac{\al^2}{c^2}\ddot{\mvec}^2-\frac{\al}{3c^2}\ddot{p}_k\eps_{kij}\tdot{\msfs}_{ij} +\frac{2}{45c^2}\ddot{p}_i\dot{\tdot{\psfs}}_{qqi}+\frac{1}{20c^2}\left(\tdot{\psfs}_{ij}\tdot{\psfs}_{ij}-\frac{1}{3}\tdot{\psfs}^2_{qq}\right)\right).
\eeqa
The output  can be partially compared with a well-known result from literature (see Refs. \cite{Landau} and   \cite{Jackson}), but for this  we have to introduce the {\it irreducible} electric and magnetic momenta defined as  symmetric trace-free (``{\bf STF}'') Cartesian tensors. Let us consider a $n-th$ order tensor $\Tsf^{(n)}$ and the corresponding projections  ${\mathcal S}(\Tsf^{(n)})$ and ${\mathcal T}(\Tsf^{(n)})$ on the subspaces of symmetric and 
{\bf STF} tensors. For  the  case of the electric moment $\psft^{(n)}$, this is a symmetric tensor and one has just to  establish  their {\bf STF} projection. Let us consider the simplest case of the quadrupolar electric moment $\psft^{(2)}$. Writing the components $\psfs_{ij}$ as
\beqan
\psfs_{ij}=\Pi_{ij}+\la \delta_{ij}\ ,
\eeqan
there is a unique value  of the parameter $\la$ such that $\Pi^{(2)}={\mathcal T}(\psft^{(2)})$. For $\la=\psfs_{qq}/3$ ,
  \beqa\label{la2}
\Pi_{ij}=\psfs_{ij}-\frac{1}{3}\psfs_{qq}\delta_{ij}=\intl_{\dom}\big(x_ix_j-\frac{1}{3}r^2\delta_{ij}\big)\,\rho\,\rmd^3x\ .
\eeqa
In equation \eref{TRP})  the octupolar electric moment $\psft^{(3)}$ is present. The {\bf STF} projection can be calculated searching the first order tensor ${\bf{\sf \Lambda}}^{(1)}$ such that the {\bf STF} projection $\bf{\sf \Pi}^{(3)}={\mathcal T}\big(\psft^{(3)}\big)$ is given by the components
\beqa\label{sfpi3}
{\sf \Pi}_{ijk}=\psfs_{ijk}-\delta_{\{ij}{\sf \Lambda}_{k\}}\ .
\eeqa
From the condition of vanishing traces of the tensor $\bf{\sf \Pi}^{(3)}$, one easily obtains:
\beqa\label{lai}
\lasf_i=\frac{1}{5}\psf_{qqi}=\frac{1}{5}\intl_{\dom}r^2\,x_i\,\rho\,\rmd^3x\ .
\eeqa
Concerning the magnetic quadrupolar  moment $\msft^{(2)}$, we have a simple procedure for {\bf STF} projection. Let us write the identity
\beqan
\msfs_{ij}=\frac{1}{2}\left(\msfs_{ij}+\msfs_{ji}\right)+\frac{1}{2}\left(\msfs_{ij}-\msfs_{ji}\right)\ ,
\eeqan
where the first bracket represents the symmetric part of this tensor, and the second one, the antisymmetric one. The symmetric part is, for this case $(n=2)$, a {\bf STF} tensor $\gamsft^{(2)}=\Tcal(\msft^{(2)})$. Therefore, 
\beqa\label{stfG3}
\msfs_{ij}=\gamsf_{ij}+\frac{1}{2}\eps_{ijk}\Nsf_k\ ,
\eeqa
where 
\beqa\label{Nk}
\Nsf_k=\eps_{kij}\msfs_{ij}=\frac{2}{3\al}\intl_{\dom}\left(\rvec\times(\rvec\times\jvec)\right)_k \rmd^3x=
\frac{2}{3\al}\intl_{\dom}\left((\rvec\cdot\jvec)\,\rvec-r^2\,\jvec\right)_k\,\rmd^3x\ .
\eeqa
Since
\beqan
\psfs_{ij}\psfs_{ij}=
\pisf_{ij}\pisf_{ij}+\frac{1}{3}\,\psfs^2_{qq}\ ,
\eeqan
we obtain
\beqa
\tdot{\psfs}_{ij}\tdot{\psfs}_{ij}-\frac{1}{3}\,\tdot{\psfs}^2_{qq}=\,\tdot{\pisf}_{ij}\tdot{\pisf}_{ij}\ .
\eeqa
Combining the magnetic quadrupolar and the electric octupolar terms from equation \eref{TRP}), with their expressions from  equations \eref{la2}), \eref{sfpi3}), and \eref{stfG3}), we  get
\beqan
-\frac{\al}{3c^2}\ddot{p}_k\eps_{kij}\tdot{\msfs}_{ij} +\frac{2}{45c^2}\ddot{p}_i\dot{\tdot{\psfs}}_{qqi}=
\frac{4}{3c^2}\ddot{p}_k\left(-\frac{\al}{4}\tdot{\Nsf}_k+\frac{1}{6}\dot{\tdot{\lasf}}_k\right)=-\frac{4}{3c^2}\ddot{\pvec}\,\cdot\tdot{\bsy{t}}
\eeqan
a consequence of the traceless character of $\pisft^{(2)},\;\pisft^{(3)}$ and $\gamsft^{(2)}$.
Here,  the vector 
\beqa\label{tdip}
\bsy{t}=\frac{\al}{4}\Nsft-\frac{1}{6}\dot{\lasft}=\frac{1}{10}\intl_{\dom}\left((\rvec\cdot\jvec)\,\rvec-2r^2\,\jvec\right)\,\rmd^3x
\eeqa
is introduced.
The last expression is obtained from equations \eref{Nk}), \eref{lai}) and applying the continuity equation together with an operation of partial integration. This is the so-called {\it electric toroidal dipole moment}~~~\cite{Dubovik-FEC}, \cite{Dubovik-rep}. We can write the radiation intensity ${\mathcal I}_{(4)}$ in terms of {\bf STF} projections of electromagnetic momenta:
\beqa\label{I(4)}
{\mathcal I}_{(4)}=\frac{1}{4\pi\eps_0c^3}\left(\frac{2}{3}\ddot{\pvec}^2+\frac{2}{3}\frac{\al^2}{c^2}\ddot{\mvec}^2+\frac{1}{20c^2}\tdot{\pisf}_{ij}\tdot{\pisf}_{ij}-\frac{4}{3c^2}\ddot{\pvec}\cdot\tdot{\bsy{t}}     \right)\ .
\eeqa
Now, we can compare this last result with the one given by equation (71,5) from   Ref. \cite{Landau} where  the term corresponding to the quadrupolar electric moment is written in terms of the {\bf STF} tensor $\bf{\sf D}^{(2)}$ defined by the components $\sf{D}_{ij}=3\pisf_{ij}$. We see that the expression given by our equation \eref{I(4)}) differs from the equation in Ref. \cite{Landau} by the term $-(4/3c^2)\ddot{\pvec}\cdot\tdot{\bsy{t}}$, i.e.\ the contribution of the toroidal electric moment. In Ref. \cite{Landau}, one calculates the contribution to the radiation of electric and magnetic dipolar  and  electric quadrupolar momenta. For this calculation, one uses the {\bf STF} tensor $\bf{\sf D}^{(2)}$ based on the invariance of the field to the substitution $\psfs_{ij}\to \sf{D}_{ij}$ or, equivalently, the gauge transformation of potentials for such a substitution. We point out that this invariance is a singular case and, for higher orders of the multipolar expansion it is not true. The toroidal electric moment does not appear because  the orders of magnitude of the different terms in the multipolar expansion are not consequently evaluated. The  same omission is done in Ref. \cite{Jackson}. In Refs. \cite{Dubovik-FEC},\cite{Dubovik-rep}, \cite{RV}, the toroidal  dipole  is obtained as a first term from a class of toroidal multipoles. The class is pointed out in a complete multipole analysis, appearing as additional terms beside the {\bf STF} projections of the primitive moments $\psft$ and $\msft$. In Ref. \cite{Zeldovich}, our formula \eref{I(4)}) is obtained as a correction to the result of Ref. \cite{Landau} in order to assure  a quantity origin independent, i.e.\ translation invariant.
The toroidal contribution from equation \eref{I(4)}) is obtained in Ref. \cite{Bellotti}, too,  by a calculation close to  the procedure from the present paper.
\par  Let us consider the contribution of the electric quadrupolar moment which, as seen from equation \eref{Brad1}), is proportional to the third  time derivative of the vector
\beqa\label{V}
\bsy{V}=\nuvec\times \bsy{\mathcal P}=\frac{1}{r}\evec_i\eps_{ijk}\nu_j\nu_q\psfs_{qk}.
\eeqa
The substitution 
\beqa\label{sbt}
\psft^{(2)}\to \pisft^{(2)}=\Tcal(\psft^{(2)})
\eeqa
 in equation \eref{V}) gives 
\beqan
\bsy{V}\to \bsy{V}-\frac{1}{3r}\evec_i\eps_{ijk}\nu_j\nu_q\delta_{qk}\psfs_{ll}=\bsy{V}-\frac{1}{3r}\psfs_{ll}(\nuvec\times\nuvec)=\bsy{V}.
\eeqan
One can see that $\Bvec(\rvec,t)$ and $\Evec(\rvec,t)$ given by equations \eref{B5}) and \eref{E1}) are also invariant if one performs the substitution \eref{sbt}). It is an exercise for the reader to prove that the substitution \eref{sbt}) in the expansions of the potentials $\Avec$ and $\Phi$ given by equations \eref{Adezv}) and \eref{Phi}) has as effect a gauge transformation of these potentials \cite{cvJPA}
\par Unfortunately, this type of invariance is not valid for magnetic moments and, for $n\ge3$, for electric ones. In these cases, the physical results are not invariant with respect to the transformations 
\beqa\label{TPMn}
\psft^{(n)}\to \Tcal(\psft^{n)}),\;\;\;\;\msft^{(n)}\to \Tcal(\msft^{n)}).
\eeqa
Let us suppose however that, at least for a given  pair of numbers $(M,N)$, all the magnetic and electric moments for $m\le M,\,n\le N$ in the expression \eref{I1}) of the radiation intensity can be substituted by {\bf STF} tensors $\msftr^{(m)},\;\psftr^{(n)}$. These new tensors do not necessarily coincide with the corresponding projections $\Tcal(\msft^{(m)}),\,\Tcal(\psft^{(n)})$ as it will be seen in the following. In this situation, the calculation of the terms from equation \eref{I1}) obtained by the corresponding substitutions $\bmu(m)\to \widetilde{\bmu}(m),\;\bpi(n)\to \widetilde{\bpi}(n)$ is more easily performed. Indeed, if $\Asft^{(n)},\;\Bsft^{(m)}$ are {\bf STF} tensors, then \cite{cvdn-JPA} 
\beqa\label{stf1}
\left\langle\big (\nu_{i_1},\dots\nu_{i_k}\Asf_{i_1,\dots i_n}\big)\big (\nu_{i_1},\dots \nu_{i_{k'}}\Bsf_{i_1,\dots i_m}\big)\right\rangle=\frac{k!\,\delta_{kk'}}{(2k+1)!!}\Asf_{i_1\dots i_n}\Bsf_{i_1\dots i_m}\equiv\frac{k!\,\delta_{kk'}}{(2k+1)!!}\left(\Asft^{(n)}\ct\Bsft^{(m)}\right).\nonumber\\
~
\eeqa
Here, since $\Asft^{(n)}$ and $\Bsft^{(m)}$ are symmetric tensors, it is of no importance  what is the order of factors in the contraction $\Asft^{(n)}\ct\Bsft^{(m)}$ ( noted  as  $\Asft^{(n)}\vert\vert\Bsft^{(m)}$, too). Obviously, the result of a  such contraction is a {\bf STF} tensor of  $\vert n-m\vert-th$ order. All the terms in equation \eref{I1}) can be calculated using equation \eref{stf1}) except for those  containing vectorial products which vanish as consequence of the equation \eref{numed}) and of the symmetry properties of the tensors $\psft^{(n)},\;\msft^{(m)}$. Let us apply equation \eref{stf1}) to the term :
\beqa\label{term1}
\left\langle \widetilde{\bmu}(t;\nuvec,n)\cdot \widetilde{\bmu}(t;\nuvec,m)\right\rangle&=&\left\langle\nu_{i_1}\dots
\nu_{i_{n-1}}\,\nu_{j_1}\dots\nu_{j_{m-1}} \right\rangle\widetilde{\msfs}_{i_1\dots i_{n-1}\,i}(t)\, \widetilde{\msfs}_{j_1\dots j_{m-1}\,i}(t)\nonumber\\
&=&\frac{(n-1)!\,\delta_{nm}}{(2n-1)!!}\left(\widetilde{\msft}^{(n)}(t)\ct\widetilde{\msft}^{(m)}\right),
\eeqa
and
\beqa\label{term2}
\left\langle \big(\nuvec\cdot\widetilde{\bmu}(t;\nuvec,n)\big)\big(\nuvec\cdot\widetilde{\bmu}(t;\nuvec,m)\big)\right\rangle=
\frac{n!\,\delta_{nm}}{(2n+1)!!}\left(\widetilde{\msft}^{(n)}\ct\widetilde{\msft}^{(m)}\right)\ .
\eeqa
 The expressions for alike terms are similar. The terms including vectorial products as, for example:
\beqa\label{term3}
\left\langle\widetilde{\bmu}(t;\nuvec,n)\cdot\big(\nuvec\times\widetilde{\bpi}(t;\nuvec,m)\big)\right\rangle
=\left\langle\widetilde{\mu}_i(t;\nuvec,n)\eps_{ijk}\nu_j\widetilde{\pi}_k(t;\nuvec,m)\right\rangle\nonumber\\
=\eps_{ijk}\left\langle\nu_{i_1}\dots\nu_{i_{n-1}}\,\nu_j\,\nu_{j_1}\dots\nu_{j_{n-1}}\right\rangle
\tilde{\msfs}_{i_1\dots i_{n-1}\,i}\widetilde{\psfs}_{j_1\dots j_{m-1}\,k}=0
\eeqa
vanish, since each term from the average  
 $\left\langle\nu_{i_1}\dots\nu_{i_{n-1}}\,\nu_j\,\nu_{j_1}\dots\nu_{j_{n-1}}\right\rangle$ given by equation \eref{numed}) includes either  $\delta_{ji_q},\,q=1\dots n-1$ or $\delta_{ji_q},\,q=1\dots m-1$ or directly an odd number of $\nu$.
 \par Using these results, we obtain the final expression of the (total) radiated power:
 \beqa\label{tpr}
 \!\!\!{\mathcal I}(t)=\frac{\al^2}{4\pi\eps_0\,c^3}\suml_{n \ge 1}\frac{n+1}{n\,n!\,(2n+1)!!\,c^{2n}}\left\{\left(\widetilde{\msft}^{(n)}_{,n+1}(t)\ct\widetilde{\msft}^{(n)}_{,n+1}(t)\right)\,+\,\frac{c^2}{\al^2}\left(\widetilde{\psft}^{(n)}_{,n+1}(t)\ct\widetilde{\psft}^{(n)}_{,n+1}(t)\right)\right\}
 \eeqa
where we consider the limit for $M,N\,\to \infty$. We point out that considering the infinite sum in equation \eref{tpr}), makes the tensors $\widetilde{\msft}^{(n)},\;\widetilde{\psft}^{(n)}$  infinite series. This is not so catastrophic since for practically applications only a finite number of terms from equation \eref{tpr}) is necessary,   as it will be seen in the following.
Obviously, one can apply equation \eref{ompm}) for the {\bf STF}-projections of the multipole momenta.
   \par For an answer  to the question how one can use equation  \eref{tpr}) for calculating the radiation intensity, let us return to the problem of calculating this quantity up to the $4-th$ order in the parameter $\zeta$. Since, together with $\psft^{(2)}$, the momenta $\msft^{(2)}$ and $\psft^{(3)}$ give contributions in this order, we have to consider in equation \eref{tpr})   the {\bf STF} projections $\gamsft^{(2)}$ and $\pisft^{(3)}$, too. This time the field is not invariant for the substitutions 
\beqa\label{sbt23}   
   \msft^{(2)}\to\gamsft^{(2)},\;\;\;\;\;\psft^{(3)}\to \pisft^{(3)}.  
   \eeqa
   \par Let us calculate the effect of the first substitution from equation \eref{sbt23}) in equation \eref{Brad1}) for expressing the radiated magnetic field. The terms  from the expansion series of $\Bvec$ affected by the substitution are 
 \beqan
&~&\frac{1}{2c^3}\frac{\d^3}{\d t^3}\left\{\big(\nuvec\cdot{\mathcal M}(\rvec,t_0;\nuvec,2)\big)\,\nuvec-{\mathcal M}(\rvec,t_0;\nuvec,2)\right\}
=\frac{1}{2c^3}\frac{\d^3}{\d t^3}\frac{1}{r}\left\{\evec_i\nu_i\nu_j\nu_l\,\msfs_{lj}-\evec_i\nu_j\,\msfs_{ji}\right\}\nonumber\\
&~&\stackrel{\msft^{(2)}\to\gamsft^{(2)}}{\longrightarrow}\frac{1}{2c^3}\frac{\d^3}{\d t^3}\frac{1}{r}
\left\{\evec_i\nu_i\nu_j\nu_l\,\msfs_{lj} -\frac{1}{2}\evec_i\nu_i\nu_j\nu_l\eps_{ljk}\Nsf_k -\evec_i\nu_j\msfs_{ji}+\frac{1}{2}\evec_i\nu_j\eps_{jik}\Nsf_k \right\}\nonumber\\
&~&=\frac{1}{2c^3}\frac{\d^3}{\d t^3}\left\{\big(\nuvec\cdot{\mathcal M}(\rvec,t_0;\nuvec,2)\big)\nuvec-{\mathcal M}(\rvec,t_0;\nuvec,2)\right\}-\frac{1}{4c^3}\frac{1}{r}\big(\nuvec\times\tdot{\bsy{N}}\big),
\eeqan
 where $\bsy{N}=\evec_i\Nsf_i=\Nsft^{(1)}$. From equation \eref{Brad1}) we can see that the alteration of $\Bvec$ by the  substitution $\msft^{(2)}\to\gamsft^{(2)}$ is compensated by the following transformation of the dipolar electric moment:
 \beqa\label{modp1}
 \pvec\to \pvec-\frac{\al}{4c^2}\dot{\bsy{N}}.
 \eeqa
Let us consider the substitution $\psft^{(3)}\to \pisft^{(3)}$ in equation \eref{Brad1}). The term affected is:
\beqan
&~&-\frac{1}{6\al c^3}\frac{\d^4}{\d t^4}\left(\nuvec\times{\mathcal P}(\rvec,t_0;\nuvec,3)\right)=-\frac{1}{6\al c^3}\frac{1}{r}\evec_i\eps_{ijk}\nu_j\nu_q\nu_l\frac{\d^4}{\d t^4}\psfs_{qlk} \nonumber\\
&~& \stackrel{\psft^{(3)}\to \pisft^{(3)}}{\longrightarrow}
-\frac{1}{6\al c^3}\frac{1}{r}\frac{\d^4}{\d t^4}\left(\eps_{ijk}\nu_j\nu_q\nu_l\,\psfs_{qlk}-\eps_{ijk}\nu_j\nu_q\nu_l\delta_{\{ql}\lasf_{k\}}\right)\nonumber\\
&~&=-\frac{1}{6\al c^3}\frac{\d^4}{\d t^4}\left(\nuvec\times{\mathcal P}(\rvec,t_0;\nuvec,3)\right)+\frac{1}{6\al c^3}\frac{1}{r}\left(\nuvec\times\dot{\tdot{\bsy{\Lambda}}}\right)\ ,
\eeqan
where $\bsy{\Lambda}=\evec_i\lasf_i$.
The alteration of $\Bvec$ by the substitution $\psft^{(3)}\to \pisft^{(3)}$ is compensated by the following transformation of the dipolar electric moment:
\beqa\label{modp2}
\pvec\to \pvec+\frac{1}{6c^2}\ddot{\bsy{\Lambda}}.
\eeqa
The total  alteration of $\Bvec$ by the two substitutions \eref{sbt23}) is compensated by the transformation
\beqa\label{pfin1}
\pvec\to \widetilde {\pvec}=\pvec-\frac{1}{c^2}\left(\frac{\al}{4}\dot{\bsy{N}}-\frac{1}{6}\ddot{\bsy{\Lambda}}\right)=\pvec-\frac{1}{c^2}\dot{\bsy{t}}\ ,
\eeqa
with $\bsy{t}$ defined by equation \eref{tdip}).
\par Now, in equation \eref{tpr}) we have to consider $\widetilde{\psft}^{(1)}=\widetilde{\pvec}$ given by equation \eref{pfin1}),   $\widetilde{\msft}^{(1)}=\mvec,\;\widetilde{\psft}^{(2)}=\pisft^{(2)},\;\widetilde{\msft}^{(2)}=\gamsft^{(2)},\;\widetilde{\psft}^{(3)}=\pisft^{(3)}$ and, for $n\ge 4$, the primitive momenta which contribute only with terms of orders greater than $4$ in $\zeta$. We can write:
\beqan
{\mathcal I}(t)=\frac{\al^2}{4\pi\eps_0c^3}\left\{\frac{2}{3c^2}\left(\ddot{\mvec}^2+\frac{c^2}{\al^2}\big(\ddot{\pvec}-\frac{1}{c^2}\tdot{\bsy{t}}\big)^2\right)+\frac{1}{20c^2\al^2}\tdot{\pisft}^{(2)}\ct \tdot{\pisft}^{(2)} \right\}+\dots\ .
\eeqan
The quadrupolar magnetic and octupolar electric terms are not written because these are of  order $6$ in $\zeta$.
Since $\tdot{\bsy{t}}^2\sim \zeta^6$, from the last expression  of ${\mathcal I}$ we are left with equation \eref{I(4)}).
\par This calculation offers a suitable example for understanding the general scheme of tensor reduction for electric and magnetic moments, as well as for the usage of formula \eref{tpr}) when describing the radiated power.  Replacing an electric moment of order $n$ by its {\bf STF}-projection induces in the tensors of inferior order, compensating terms of order $\zeta^n$. The general term in the expansion \eref{tpr}) is of order $\zeta^{2n}$. If we are interested in the approximation of ${\mathcal I}$ up to terms of order $\zeta^{2k}$, then we have to replace the tensors related to the moments $\psft^{(2k-1)}$ and $\msft^{(2k-2)}$  by {\bf STF}-projections and to take into account all compensating terms for the lower-order tensors.  

\section{Some basic formulae for generalizing the procedure of reducing the electric and magnetic momentum tensors}
\label{sec:formulae}
In this section we list the required formulae for the generalized tensor reducing procedure. The equations are given without the related proofs since they can be found in the literature.

\par Let $\Ssft^{(n)}$ be a symmetric tensor of rank  $n$. Its {\bf STF}- projection results from the following formula: 
\beqa\label{TSn}
\Tcal\big(\Ssft^{(n)}\big)_{i_1\dots i_n}=\Ssft_{i_1\dots i_n}-\delta_{\{i_1i_2}\lasf(\Ssft^{(n)})_{i_3\dots i_n\}}.
\eeqa
The operator  $\lasft$ is defined on the account of a formula for the  {\bf STF}-projection  of a symmetric tensor given in  \cite{Thorne} (the book \cite{Pirani} is cited as the origin of this formula):
\beqa\label{Lambda}
\lasf\big(\Ssft^{(n)}\big)_{i_1\dots i_{n-2}}=\suml^{[n/2-1]}_{m=0}\frac{(-1)^m[2n-1-2(m+1)]!!}{(m+1)(2n-1)!!}\delta_{\{ i_1i_2}\dots\delta_{i_{2m-1}i_{2m}}
 \Ssf^{(n;m+1)}_{i_{2m+1}\dots i_{n-2}\}}.
\eeqa
The proof can be found in \cite{Apple}.  $[a]$ stands for the integer part of $a$ and the notation  $\Ssf^{(n;p)}$  indicates $p$ pairs of contracted indices.  In the following, for simplifying the notation, all arguments of the operator $\lasft$ should be considered as  symmetric tensor i.e.\ $\lasft(\Tsft^{(n)})=\lasft(\Scal(\Tsft^{(n)})$ for any tensor $\Tsft^{n)}$. The same applies to the operator $\Tcal$: $\Tcal(\Tsft^{(n)})=\Tcal(\Scal(\Tsft^{(n)})$, $\Scal$ being the symmetrization operator.
\par In the symmetrization  process we have to calculate the symmetric projections of some tensors $\Lsft^{(n)}$ of the magnetic moment type: they are  symmetric in the first $n-1$ indices and the contraction of $i_n$ with any index $i_q,\;q=1\dots n-1$ gives a null result. For the symmetric projection of such a tensor, we introduce the formula:
\beqa\label{simL}
\big(\Scal(\Lsft^{(n)})\big)_{i_1\dots i_n}=\Lsf_{i_1\dots i_n}-\frac{1}{n}\suml^{n-1}_{\la=1}\eps_{i_{\la}i_nq}\Ncal^{(\la)}_{i_1\dots i_{n-1}q}(\Lsft^{(n)}),
\eeqa
where $\Ncal^{(\la)}_{i_1\dots i_{n-1}}$ is the component with the $i_{\la}$ index  suppressed. The operator $\Ncal$ defines a correspondence between $\Lsft^{(n)}$ and a tensor of rank $(n-1)$ of the same type from  the symmetry point of view. Particularly,
\beqa\label{N(M)}
\Ncal^{2k}\big(\msft^{(n)}\big)&=&\frac{(-1)^k n}{(n+1)\al}\intl_{\dom}(r^2)^k\,\rvec^{n-2k}\times\jvec\,\rmd^3x,\nonumber\\
\Ncal^{2k+1}\big(\msft^{(n)}\big)&=&\frac{(-1)^k n}{(n+1)\al}\intl_{\dom}(r^2)^k\,\rvec^{n-2k-1}\times(\rvec\times\jvec)\,\rmd^3x,\;\;k=0,\,1,\,2,\dots
\eeqa
where $\bsy{a}^n\times\bsy{b}$ is the tensor defined by the components $\big(\bsy{a}^n\times\bsy{b}\big)_{i_1\dots i_n}=a_{i_1}\dots a_{i_{n-1}}(\bsy{a}\times\bsy{b})_{i_n}$.
Let us consider the process of {\bf STF} reducing the multipole tensors starting with  the rank $N+1$  in the electric case, and  $N$ in the magnetic one. In Ref. \cite{CV-lett} one finds  a general formula for the resulting tensors $\widetilde{\psft}^{(n)}$, for $n=1,\,\dots,\,N$ and  $\widetilde{\msft}^{(n)}$, for $n=1,\,\dots, \,N-1$. It includes  all compensating terms obtained in this process
\beqa\label{Pgen}
\widetilde{\psft}^{(n)}&=&\Tcal\big({\psft}^{(n)}\big)+\suml^{[(N-n)/2]}_{k=1}\frac{(-1)^k}{c^{2k}}\,\frac{\d^{2k-1}}{\d t^{2k-1}}\,\Tsft^{(n)}_k,\nonumber\\
\Tsft^{(n)}_k&=&(-1)^kc^{2k}\,\Tcal\left(A^{(n)}_k\lasf^k\big(\dot{\psft}^{(n+2k)}\big)+\suml^{k-1}_{l=0}B^{(n)}_{k-1,\,l}\lasf^l\,\Ncal^{2k-2l-1}\big(\msft^{(n+2k-1)}\big)\right),
\eeqa
and
\beqa\label{Mgen}
\widetilde{\msft}^{(n)}=\Tcal\big(\msft^{(n)}\big)+\Tcal\left(\suml^{[(N-n-1)/2]}_{k=1}\frac{\d^{2k}}{\d t^{2k}}\suml^k_{l=0}C^{(n)}_{kl}\lasf^l\Ncal^{2k-2l}\big(\msft^{(n+2k)}\big)\right).
\eeqa
The coefficients are given in a compact algebraic form in Ref. \cite{cvcs}:
\beqa\label{ABC}
A^{(n)}_k&=&\frac{1}{2^k\,c^{2k}}\,\frac{n}{n+2k},\nonumber\\
B^{(n)}_{k,l}&=&\frac{(-1)^{k-l+1}\al}{2^l\,c^{2k+2}}\,\frac{n(n+2l)!}{(n+2k+1)(n+2k+1)!},\nonumber\\
C^{(n)}_{k,l}&=&\frac{(-1)^{k-l}}{2^l\,c^{2k}}\frac{n(n+2l)!}{(n+2k)(n+2k)!}.
\eeqa
The reader is encouraged to apply these formulae for the cases $N=4,\,5,\,6$ and to find convenient intermediary  calculation. We point out that in equation \eref{Pgen}) the normalisation of the quantities $\Tsft^{(n)}$ is chosen such that the quantities $\Tsft^{(n)}_1$ coincide with the electric toroidal moments given in literature at least for $n$ up to $3$.

%######################################################
\section{Conclusion and discussion}\label{sec:concl}
\par After an introduction on the formalism for multipolar expansions of the electric and magnetic field, we dedicated sections \ref{sec:radiation} and \ref{sec:jefimenko} to the main purpose of this article: the analysis of different methods for calculating the radiated field with an emphasis on the novelty the use of Jefimenko's equations can bring. In the last two sections we presented aspects of the radiated power calculations from the tensorial point of view. 
\par From the analysis of sections \ref{sec:radiation} and \ref{sec:jefimenko} we can draw some conclusions on the utility of Jefimenko's equations when considering the multipolar expansion problem. As one can notice, when deriving equation \eref{B4-0}) from \eref{B3}), no matter if we work with potentials or directly with fields, the inversion of the spatial derivative with the integral on the domain $\dom$ is mandatory. 
%For the radiation field, as it can be seen when deriving equation \eref{B9}), such an inversion of operations is not %necessary.
In Section \ref{sec:jefimenko} it was proven that the multipole expansion of the fields $\Evec$ and $\Bvec$ can be obtained generally and directly from Jefimenko's equations. All expanding operations are performed on the integrand, but, still under the assumption that the integration operation is distributive with respect to the Taylor series. It remains only to argue the necessity of the corresponding additional calculation effort for applying this procedure. 
\par As an additional remark, we emphasize that in the same sections we tried to remind the reader that if one wants to completely describe the radiative systems, one has to include besides terms of order $1/r$, the $1/r^2$  contributions from the field expansions. 
\par Ref. \cite{Melo} is part of a paper series trying to emphasize the theoretical and practical importance of Jefimenko's equations. These equations are considered as {\it extraordinarily powerful and illuminating} as the authors of Ref. \cite{Griff} write. We have nothing against the open enthusiasm in these papers. We neither dispute the beauty of the result regarding the calculation of the retarded fields $\Evec$ and $\Bvec$ directly from Maxwell's equations, without having to introduce and handle the potentials. Maybe a series of applications based on these equations are more efficacious and physically more transparent. Although, we are circumspect concerning the axiomatic treatment of the electromagnetic theory starting from these equations (opposite to opinions from e.g. Refs. \cite{Jefimenko}, \cite{Heras}), but,  this subject  will be discussed elsewhere.
\par In the second part of the paper we gave a detailed calculation for the radiated power up to the $4-th$ order in $d/\la\,<\,1$. The method can be similarly applied to higher-orders.  We related our procedure to existing prescriptions in the literature and we underlined what omissions might show when particular terms of the multipolar expansions are neglected. We hope we were able to convince the reader how powerful and not so complicated a consistent vectorial/tensorial computation is.
\vspace{1cm}
\begin{acknowledgments}
The work of RZ was supported by grant ID946 (no.\ 44/2007) of Romanian National Authority for Scientific Research.
\end{acknowledgments}

\end{document}